\documentclass[a4paper]{jpconf}
\usepackage{graphicx}
\usepackage{lineno}
\usepackage{hyperref}

\begin{document}

\title{Initial explorations of ARM processors for scientific computing}

\author{David Abdurachmanov$^1$, Peter Elmer$^2$, Giulio Eulisse$^3$, Shahzad Muzaffar$^3$}

\address{$^1$ Digital Science and Computing Center, Faculty of Mathematics and Informatics, Vilnius University, Vilnius, Lithuania}
\address{$^2$ Department of Physics, Princeton University, Princeton, NJ 08540, USA}
\address{$^3$ Fermilab, Batavia, IL 60510, USA}

\ead{Peter.Elmer@cern.ch}

\begin{abstract}
Power efficiency is becoming an ever more important metric for both high
performance and high throughput computing. Over the course of next decade 
it is expected that flops/watt will be a major driver for the evolution of 
computer architecture. Servers with large numbers of ARM processors, already
ubiquitous in mobile computing, are a promising alternative to
traditional x86-64 computing. We present the results of our initial 
investigations into the use of ARM processors for scientific computing 
applications. In particular we report the results from our work
with a current generation ARMv7 development board to explore ARM-specific 
issues regarding the software development environment, operating system, 
performance benchmarks and issues for porting High Energy Physics software.
\end{abstract}

\section{Introduction}

The computing requirements for high energy physics (HEP) projects like the 
Large Hadron Collider (LHC)~\cite{LHCMACHINE} at the European
Laboratory for Particle Physics (CERN) in Geneva, Switzerland are
larger than can be met with resources deployed in a single computing
center. This has led to the construction of a global distributed computing 
system known as the Worldwide LHC Computing Grid (WLCG)~\cite{WLCG},
which brings together resources from nearly 160 computer centers in 35 
countries. Computing at this scale has been used, for example, 
by the CMS~\cite{CMSDET} and ATLAS~\cite{ATLASDET} experiments for the 
discovery of the Higgs boson~\cite{CMSHIGGS,ATLASHIGGS}. To achieve this
and other results the CMS experiment, for example, typically used during 
2012 a processing capacity between 80,000 and 100,000 x86-64 cores from
the WLCG.
Further discoveries are possible in the next decade as the LHC moves to 
its design energy and increases the machine luminosity. However, increases in
dataset sizes by 2-3 orders of magnitude (and commensurate processing
capacity) will eventually be required to realize the full potential of this
scientific instrument. The scale and longevity of the LHC computing 
require continual R\&D into new technologies which may be relevant in
the coming years. In this paper we report on our investigations into
one such technology, low power ARM processors, for scientific computing.

\section{Processor Architectures and Power Efficiency}

The construction of the WLCG was greatly facilitated by the convergence
around the year 2000 on commodity x86 hardware and the standardized
use of Linux as the operating system for scientific computing clusters. 
Even if multiple generations of x86 hardware (and hardware from both Intel 
and AMD) are provided in the various computer centers, this was a far
simpler situation than the typical mix of proprietary UNIX operating
systems and processors. 

Until around 2005, a combination of increased 
instruction level parallelism and (in particular) processor clock frequency 
increases insured that performance gains expected from Moore's Law would 
be seen by single sequential applications running on a single processor.
The combination of Linux, commodity x86 processors and Moore's law gains
for sequential applications made for a simple software environment.

Since around 2005, however, processors have hit scaling limits, largely
driven by overall power consumption~\cite{GAMEOVER}. The first large 
change in commercial processor products as a result of these limits
was the introduction of ``multicore'' CPUs,
with more than one functional processor on a chip. At the same time
clock frequencies ceased to increase with each processor generation and 
indeed were often reduced relative to the peak. The result of this was
that one could no longer expect that single, sequential applications would
run faster on newer processors. However in the first approximation,
the individual cores in the multicore CPUs appeared more or less
like the single standalone processors used previously. Most large
scientific applications (HPC/parallel or high throughput) run in
any case on clusters and the additional cores are often simply
scheduled as if they were additional nodes in the cluster. This
allows overall throughput to continue to scale even if that of a
single application does not. It has several disadvantages, though,
in that a number of things that would have been roughly constant
over subsequent purchasing generations in a given cluster (with
a more or less fixed number of rack slots, say) now grow with each
generation of machines in the computer center. This includes the
total memory required in each box, the number of open files and/or
database connections, increasing number of independent (and incoherent)
I/O streams, the number of jobs handled by batch schedulers,
etc.  The specifics vary from application to application, but
potential difficulties in continually scaling these system parameters
puts some pressure on applications to make code changes in response,
for example by introducing thread-level parallelism where it did
not previously exist.

There is moreover a more general expectation that the limit of power
consumption on future Moore's Law scaling will lead to more profound
changes going forward. In particular, the power hungry x86-64
``large'' cores of today will likely be replaced by simpler and less power 
hungry ``small'' cores with a greater emphasis on aggregate throughput
performance per watt, rather than just raw performance. This has 
rekindled interest in solutions that would lead back to a more
heterogeneous computing environment.

\section{ARM Processors}

A strong contender for this evolving low power (high performance/watt)
server market is the ARM processor~\cite{ARMPROC} due its nearly
complete dominance
in the low power mobile market for smartphones and tablets, which has
also seen dramatic growth since around 2005. The size of the mobile
market, and its traditional focus on low power,
has led to interest in using these processors also in a server
environment. As such ARM-based server products such as Boston
Viridis~\cite{VIRIDIS} are starting to appear.

The ARM processor has a long history~\cite{ARMSTORY}
dating back to Acorn Computers and the early days of
personal computers. It is a RISC processor and the current 
generation (ARMv7), used in most high-end mobile devices and the new server products, is a 32bit processor. We are interested in the ``A'' series of
general purpose ``Application'' processors. (ARM also produces
``R'' and ``M'' series designs for use in real-time and embedded microcontroller environments, respectively.)
A 64bit version of the ARM processor (ARMv8) has also been
designed and is expected to appear in server products from fall
2013 or early 2014.
Intel has also announced the development
of products (Silvermont) aimed at a low power market. However,
much like the current
mobile market, it isn't positioned to dominate the low power server market as it has dominated
the commodity processor market in the past.

As the ARM processors are general purpose and run Linux, only a 
standard port of the CMS software is required, similar to
what was done, for example, to port the CMS software from
32bit (ia32) to 64bit (x86-64). Such a port is reasonably
straightforward relative to the changes required to use other high
performance per watt
solutions (e.g. GPGPU's, which require actual software rewrites), thus the
effort required for these initial investigations was also relatively
modest.

\section{Test Setup}

For the tests described in this paper we used a low-cost development board,
the ODROID-U2~\cite{ODROID}. The processor on the board is an 
Exynos 4412 Prime, a System-on-Chip (SoC) produced by Samsung for use
in mobile devices. It is a quad-core Cortex A9 ARMv7 processor operating
at 1.7GHz with 2GB of LP-DDR2 memory. The processor also contains an ARM 
Mali-400 quad-core GPU accelerator, although that was not used for the work
described in this paper. The board has eMMC and microSD slots, two USB 2.0
ports and 10/100Mbps Ethernet with an RJ-45 port. Power is provided 
a 5V DC power adaptor.

The cost of the board alone was \$89 and with the relevant
accessories (cables, a cooling fan, a 64GB eMMC storage module, etc.)
the total cost was \$233. This extremely modest cost permitted us to do
meaningful initial investigations without investing in a full-fledged server.

All build tests were done using a 500GB $3.5\,''$ ATA disk connected via USB.
Runtime tests were done with output written to the eMMC storage.

For the Linux operating system on the board we used Fedora 18 ARM Remix
with kernel version 3.0.75 (provided by Hardkernel,
the vendor for the ODROID-U2 board) due to its similarities to
Scientific Linux CERN (SLC). It is fully hard float
capable and uses the floating point unit on the SoC. The kernel was
reconfigured to enable swap devices/files, which is required for
CMSSW compilation. A 4GB swap file was used in our build environment.

In order to compare results from the ARM board we also used two typical x86-64
servers currently deployed at CERN. The first is a dual quad-core Intel Xeon
L5520 $@$ 2.27 GHz (Nehalem) with 24GB of memory. The second is dual
hexa-core Intel Xeon E5-2630 $@$ 2.00GHz (Sandy Bridge) with 64GB of
memory. Both machines were equipped with a large local disk for
output and used software installed on an afs filesystem at CERN.
These machines were
purchased about three years apart and very roughly represent the range
of x86-64 hardware being operated at the time of our ARM tests.

\section{Experimental Results}

\subsection{The CMSSW software stack}

The software written by the CMS collaboration itself (CMSSW) consists of 
approximately 3.6M source lines of code (SLOC), as measured by the SLOCCount 
tool~\cite{SLOCCOUNT}. The entire software stack includes also 125 ``external''
packages, including HEP software packages like ROOT~\cite{ROOT} ($\sim$1.7M SLOC),
Geant4~\cite{GEANT4} ($\sim$1.2M SLOC) and many general open source
packages: GCC, boost, Qt, Python, etc.

\subsection{Build platform for ARM tests}
For the compilation and linking of this large set of software we considered
three options: compilation directly on the ODROID-U2 board itself,
cross compilation for the ODROID-U2 board from an x86-64 host and
compilation for the ODROID-U2 board from an emulation environment
such as QEMU~\cite{QEMU} running on an x86-64 host

Our experience with ARM emulation with QEMU prior to purchasing the
ODROID-U2 led us to believe that it was not yet quite mature enough to
provide a stable build environment sufficient for our needs. Even
though it was clear that the small ODROID-U2 development board is much less
powerful than most standard x86-64 servers, we decided to attempt
compilation of the full stack directly on the ARM board. This was motivated
by the idea that should we eventually adopt ARM as a standard production
architecture, we would probably aim for direct compilation on ARM
servers. 

\subsection{Issues arising during the ARM port}
During the port to ARMv7 we had to resolve a number of issues in the
software and build recipes:

{\bf Oracle:} We did not have Oracle libraries for the ARMv7 architecture. By construction, however, no standard Grid-capable CMS applications depend
on Oracle. Thus this affected only a small number of CMSSW packages
used primarily for writing calibrations into the Oracle database
at CERN. CMS applications which read calibrations do not interact
directly with Oracle, but instead access the calibrations via the
FroNTier web service~\cite{FRONTIER}, with no direct dependency on Oracle.

{\bf Configuration:} There were a number of minor configuration issues, for example:
  \begin{itemize}
   \item The {\tt -m32} and {\tt -m64} options do not work. (On x86-64
        CMS had made a complete transition to 64bit a couple of years
         prior to this work.)
   \item In a number of places there were x86-based assumptions leading
         to attempts to configure for x86-64 SSE and AVX
  \end{itemize}

{\bf Memory use:} Compilation of some translation units (primarily
generated ROOT dictionaries) exhausted the virtual memory address space.
Here the solution was simply to refactor the dictionaries.

{\bf ROOT Cintex:} A patch was needed for the Cintex trampoline in ROOT
to support the ARM architecture. This was submitted to the ROOT developers.

{\bf Signedness:} x86-64 and ARM treat the signedness of 
char/bit-fields differently, Intel is signed and ARM is unsigned by
default. This was dealt with on ARM by imposing the use of the compiler
options {\tt -fsigned-char} and {\tt -fsigned-bitfields},
along with a few small code modifications to fix non-portable code.

{\bf Dictionary generation and I/O:} There were several bugs in the
ROOT I/O infrastructure, as well as non cross-platform types (that 
crept in after CMS transitioned to 64bit on x86-64 and stopped
regularly producing 32bit builds), that at time of these tests
prevented us from properly reading and writing ROOT files.

\subsection{Build times}
With changes to the build recipes resulting from fixing these issues
we were able to build all of the standard CMSSW externals and 99\%
of the CMSSW code. The CMSSW code which did not build was the small
subset requiring Oracle. This demonstrates the advantages of relying
primarily on open source software and, when closed source software
cannot be avoided, carefully restricting the code which can depend
on the closed source libraries.

After making the changes described above, we achieved the following results
for the total build times directly on the ODROID-U2 board:
\begin{itemize}
\item $\sim$4h - compilation of a ``bootstrap'' kit consisting of the GCC compiler (version 4.8.0)
and a small set of packages (rpm, apt, zlib, ncurses, nspr, sqlite, etc.)
that we use for packaging and distributing the results of our builds
\item $\sim$12h - compilation of all of the 125 external packages not
included in the ``bootstrap'' kit
\item $\sim$25.5h - compilation of the CMSSW code itself as well as a set of
generated ROOT dictionaries
\end{itemize}
These are quite reasonable results. Taking into account that the 
externals do not change frequently, these results are already very close
to consistent with an eventual ``nightly'' integration build where we
compile the very latest versions of all of the CMSSW code, but reuse
pre-existing builds of the externals.

\subsection{Run time tests}

For a run time test and benchmark we used an actual CMS application
from the build described above rather than a synthetic benchmark. 
This application performs a
Monte Carlo simulation of 8 TeV LHC Minimum bias events using
Pythia8~\cite{PYTHIA8} (event generation)
followed by simulation with Geant4~\cite{GEANT4}.
Due to the problem mentioned above with generation of the dictionaries
used for data input/output, output was turned off. The data output 
however has little
effect on the total CPU time (and thus these benchmarks) as the CPU
cost is heavily dominated by the Geant4 simulation. The application
itself is single-threaded (sequential) and thus ran on a single core
at a time. To simplify testing we ran only a single job at a time
on each machine with the aim of measuring the single core
performance. Multiple
tests were performed of sufficient length to estimate properly
an average per event time and dedicated tests were run to subtract
off job startup times. A proper validation of the application output
was complicated by the lack of an output file, but checks
done by enabling printout indicated consistency between ARM and x86-64.
The results for performance (events simulated per minute per core) are
shown in Table~\ref{tab:results}. 

\begin{table}[ht]
\caption{Results of run time tests}
\centering
\label{tab:results}
\begin{tabular}{lcccc}
  \hline \hline
      &       &             & Events & Events \\
      &       &             & /minute& /minute \\
 Type  & Cores &Power (TDP) & /core  & /Watt  \\ \hline \hline
 Exynos4412 Prime      &     &     &      &      \\
 @ 1.704 GHz  & $4$ & 4W & 1.14 & 1.14 \\ \hline
 dual Xeon L5520     &     &     &      &      \\
 @2.27GHz       & $2\times4$ & 120W & 3.50 & 0.23 \\ \hline
 dual Xeon E5-2630L      &     &     &      &      \\
@2.0GHz  & $2\times6$ & 120W & 3.33 & 0.33 \\ \hline
\end{tabular}
\end{table}

In order to calculate values for the performance per watt, it
would be a bit misleading to compare the total power used by
real (x86-64) servers with a small development board. To get
a better estimate, which more directly compares the processors
themselves, we used the ``thermal design power'' (TDP) numbers
for the processors themselves.
Here TDP numbers for the two Intel Xeon x86-64 processors were taken
from their website~\cite{XEONTDP}.
For the ODROID-U2 we were
not able to find specific TDP numbers, but based on our own measurements
we have estimated the TDP-equivalent fully loaded power at about 4W.
Using these values we have calculated an equivalent performance (in
events per minute) per watt in the last column of Table~\ref{tab:results}.
A clear advantage in terms of performance per watt is seen for the
ARM processor.

\section{Conclusions}
We have done a port of the entire CMS software stack, including 3.6M SLOC
of CMS-written code in CMSSW and 125 external support packages, to
the ARMv7-based ODROID-U2 development board. We chose to build directly
on the development board itself and measured build times consistent
with operating a ``nightly'' build of the CMS software. We report
performance and performance per (TDP) watt numbers both for the ARMv7
board and for two typical x86-64 servers at CERN. On the basis of
these results we conclude that ARM-based low power servers, if they
succeed in the market, show great potential for use with typical HEP
high throughput computing applications.

\section*{Acknowledgment}
This work was partially supported by the National Science Foundation, under
Cooperative Agreement PHY-1120138, and by the U.S. Department of Energy.

\end{document}